\documentstyle[epsfig,aps]{revtex}
\begin{document}
\bibliographystyle{unsrt}
\newcommand{\Gen}{\mbox{G$_{E}^{n}$}}
\newcommand{\GenQ}{\mbox{G$_{E}^{n}({\mathrm{Q}}^2)$}}
\newcommand{\Genq}{\mbox{G$_{E}^{n}({\bf q}^2)$}}
\newcommand{\Gdipole}{\mbox{G$_{D}$}}
\newcommand{\Gmn}{\mbox{G$_{M}^{n}$}}
\newcommand{\Gmnq}{\mbox{G$_{M}^{n}({\bf q}^2)$}}
\newcommand{\sQsq}{\mbox{\scriptsize Q$^2$=0}}
\newcommand{\vbl}{\mbox{\large\bf $\vert$}}
\newcommand{\vQzero}{\mbox{$^{\vbl}_{\sQsq}$}}
\newcommand{\Qsq}{\mbox{Q$^2$}}
\newcommand{\Qquad}{\mbox{Q$^4$}}
\newcommand{\rms}{\mbox{$<\!r_{n}^2\!\!>$}}
\newcommand{\rquad}{\mbox{$<\!r_{n}^4\!\!>$}}
\newcommand{\fourth}{\mbox{$4^{th}$}}
\newcommand{\rdelta}{\mbox{$<\!r_{N \to \Delta}^2\!\!>$}}
\newcommand{\GCndelta}{\mbox{G$_{\mathrm{C}2}^{N \to \Delta}$}}
\newcommand{\GCndeltaq}{\mbox{G$_{\mathrm{C}2}^{N \to \Delta}$({\bf q}$^2$)}}
\newcommand{\GMndeltaq}{\mbox{G$_{\mathrm{M}1}^{N \to \Delta}$({\bf q}$^2$)}}
\newcommand{\GCndeltaQ}{\mbox{G$_{\mathrm{C}2}^{N \to \Delta}$(\Qsq)}}
\newcommand{\Qndelta}{\mbox{${\cal Q}_{N \to \Delta}$}}
\newcommand{\NDelta}{\mbox{$N$$\to$$\Delta$}}
\newcommand{\etal}{\mbox{\it et~al.}}
\newcommand{\bea}{\begin{eqnarray}}
\newcommand{\eea}{\end{eqnarray}}
\newcommand{\be}{\begin{equation}}
\newcommand{\ee}{\end{equation}}
%
\newcommand{\xbf}[1]{\mbox{\boldmath $ #1 $}}

\fnsymbol{footnote}
\title{Moments of the neutron charge form factor 
       and the $N$$\to$$\Delta$ quadrupole transition}

\author{P. Grabmayr~\cite{pgemail} and A. J. Buchmann~\cite{abemail}}
\address{Physikalisches Institut,\\
Universit\"at T\"ubingen,
D-72076 T\"ubingen, Germany }

\date{\today}
\maketitle
\begin{abstract}
  Recent data allow a new parametrization of the neutron charge form factor
  \Gen. A parameter-free quark-model relation between \Gen\ and the \NDelta\
  quadrupole form factor \GCndelta\ is used to predict \GCndelta\ from
  \Gen\ data. In particular, \rms\ is related to \NDelta\ quadrupole moment
  \Qndelta, while \rquad\ connects to the \NDelta\ quadrupole transition
  radius \rdelta.  From the latter we derive an experimental value for the
  charge radius of the light constituent quarks $r_{\gamma q}=0.8$~fm.
  Finally, the C2/M1 ratio in pion electroproduction is predicted from the
  elastic neutron form factor data.
\end{abstract}

\pacs{13.40.Gp, 13.40.Em, 14.20.Dh, 14.20.Gk, 12.39.Jh}

\begin{figure}[h]
\begin{center}
\epsfxsize 7. true cm 
\epsfbox{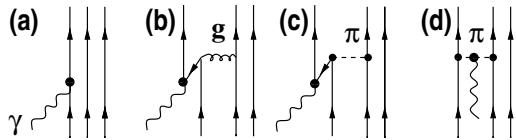}
\vspace{0.3 cm}
\caption{\label{fig:feynmec}
  Feynman diagrams of the four-vector current $J^{\mu}$
  with photon coupling to (a) one-body current
  $J^{\mu}_{[1]}$, and to (b-d) two-body gluon and pion exchange
  currents~$J^{\mu}_{[2]}$. }
\end{center}
\end{figure}

{\it Intrinsic nucleon structure -- }
The nucleon is a complicated many-particle system composed of valence quarks,
 which carry the quantum numbers, and nonvalence quark degrees of freedom,
 which describe the cloud of quark-antiquark ($q \bar q$) pairs and gluons.
 The constituent quark model~(CQM) with two-body exchange currents describes
 both these aspects of nucleon structure~\cite{Buc91}.  One-body currents
 describe the interaction of the photon with one valence quark at a time.
 Two-body exchange currents are connected with the exchange particles (gluons,
 pions) and with $q \bar q$ pairs (see Fig.~\ref{fig:feynmec}(b-d)). Nucleon
 properties which are dominated by two-body exchange currents show their
 common dynamical origin in analytical interrelations.

In this paper, a quark model relation between the neutron charge form factor
\Gen\ and the quadrupole transition form factor \GCndelta\ is used together
with a parametrization of new \Gen\ data to predict \GCndelta\ and the C2/M1
ratio usually measured by pion electroproduction.
The spin-isospin dependence of the two-body charge operator $\rho_{[2]}$, e.g.
for the gluon, can be written schematically as
\begin{equation}\label{eq:structure}%
\rho_{[2]}\approx\!\sum_{i\neq j}\!e_i\!\left[%
 \xbf{\sigma}_i\!\cdot\!\xbf{\sigma}_jY^0({\bf q})%
 -\frac{\sqrt{6}}{2}\Bigl[\![\xbf{\sigma}_i\!\times\!\xbf{\sigma}_j]^2\!\times\!
Y^2({\bf q})\!\Bigr]^0 \right]
\end{equation}
where {\bf q} is the three-momentum transfer, ${e_i}$ and $\xbf{\sigma_i}$
the quark charge and spin.  Explicit expressions can be found in 
Ref.~\cite{Buc91}.

%
As a result of (i) the dominance of $\rho_{[2]}$~\cite{Buc91,Buc97} and of
 (ii) the spin-isospin structure in $\rho_{[2]}$ and in the $N$ and $\Delta$
 wave functions, a connection between the neutron charge form factor \Genq\
 and the \NDelta\ quadrupole transition form factor \GCndeltaq\
 emerges~\cite{Buc00} as follows:
\begin{equation}\label{eq:gcgenff}
\GCndeltaq  = -{3\sqrt{2}\over {\bf q}^2} G_E^n({\bf q}^2) \ .
\end{equation}
The derivation is independent of the spatial part of the quark wave functions
 and holds irrespective of whether gluon or pion exchange or both are
 employed.  One-body contributions to \Gen\ and \GCndelta\ arise only through
 excited-state admixtures which amount to less than 20\% of the empirical
 values if a quark core radius consistent with the excitation
 spectrum~\cite{Buc97,gian} is used.  Three-body corrections are estimated to
 contribute less than 30\% of the two-body currents using a QCD
 parametrization~\cite{morp} and a large $1/N_c$ approach~\cite{lebed}.  These
 approaches also show the dominance of the two-body exchange currents for both
 observables.

The dominance of nonvalence quark degrees of freedom in \Gen\ and \GCndelta\
 is not specific to this quark model. Also chiral approaches, e.g.  the Skyrme
 model~\cite{Wir87}, the $\sigma$-model~\cite{Goli}, the Nambu-Jona-Lasinio
 model~\cite{Chr96}, and the chiral soliton model with quarks~\cite{Broni}
 predict that the neutron charge radius \rms\ and \Qndelta, the transition
 quadrupole moment, are dominated by nonvalence quark degrees of freedom.  In
 these models, the valence quark contribution to \rms\ and \Qndelta\ is
 comparatively small.  The quark model, the chiral approaches~\cite{mano}, and
 chiral perturbation theory~\cite{lebed2} have the same underlying group 
theoretical structure,
 i.e. spin-isospin symmetry.  Eq.(\ref{eq:gcgenff}) has been derived within
 the constituent quark model for the case of dominance of two-body
 operators. However, a larger range of validity is conjectured because of
 stringent constraints on allowed operator structures in
 Eq.(\ref{eq:structure}) due to the scalar nature of the charge operator and
 the spin-isospin symmetry of the $N$ and $\Delta$ wave functions.

For further interpretation of the new \Gen\ parametrization we rely on
 explicit expressions for the lowest moments of \Gen\ and \GCndelta\
 calculated in the quark model with two-body currents. In this
 model~\cite{Buc91}, the neutron charge radius can be expressed in terms of
 the empirical \mbox{$N$-$\Delta$} mass splitting and the quark core
 radius~$b$
\begin{equation}\label{eq:rnmec}
\rms = -b^2\, { M_{\Delta}-M_N \over M_N} \  .
\end{equation}
Inserting the experimental neutron charge radius
 \rms=-0.113~fm$^2$~\cite{kopecky} and the experimental $N$-$\Delta$ mass
 splitting into Eq.(\ref{eq:rnmec}) one obtains $b=0.60$~fm for the quark core
 radius, which measures the spatial extent of the valence quark wave function.
 A related quantity of interest is the slope of \GCndelta\ at ${\bf q}^2$=0,
 which is given by \cite{Buc97,Buc00}
\begin{equation}\label{eq:quadrad}
<\!r^2_{N \to \Delta}\!\!>= {11\over 20}\, b^2 +r^2_{\gamma q} \ ,
\end{equation}
where $r^2_{\gamma q}$ is the charge radius of the constituent quark.
 In the same model we find \GMndeltaq=$-\sqrt{2}$~\Gmnq.

%

{\it Moments of form factors -- }
First, at low momentum transfers $Q^2$=$-q_{\mu}^2$ any form factor,
 (here \Gen), can be expanded into a Taylor series 
\be\label{eq:genexpand}
\GenQ=\Gen(0)
          +\! \frac{\mathrm{d}\Gen}{\mathrm{d}Q^2}
\biggl {\vert}_{_{\sQsq}}\hspace*{-7,5mm}\cdot\ \Qsq
          +\!
\frac{1}{2}\frac{\mathrm{d}^2~\Gen}{(\mathrm{d}Q^2)^2}\biggl
{\vert}_{_{\sQsq}}\hspace*{-7,5mm}\cdot\ \Qquad
          +\cdots
\ee
Second, in the Breit-frame (\Qsq=${\bf q}^2$) the form factor \Gen\ is
 related to the spatial charge distribution $\rho(r)$ through the Fourier
 transform as
\be\label{eq:trans}
\GenQ = \!\int_{\mathrm{o}}^\infty\!\rho(r)e^{-i {\bf q\cdot r}}
\mathrm{d}{\bf r}
 = 4\pi\!\!\int_{\mathrm{o}}^\infty\!\rho(r)\frac{\sin{qr}}{qr}
r^2\mathrm{d}r \ .
\ee
A series expansion of $\sin(\mathrm{qr})$ leads to the well known relation
 between the derivatives of the electric form factor and
 the radial moments of the charge distribution
\be\label{eq:genderivone}
\Gen'(0)  \equiv 
\frac{\mathrm{d}\Gen}{\mathrm{d}Q^2}\biggl {\vert}_{_{\sQsq}}\hspace*{-6mm}
= -\frac{\rms}{6}  = -\frac{a\mu_n }{4M_n^2} \ ,
\ee
\be\label{eq:genderivtwo}
\Gen''(0)\equiv%
\frac{\mathrm{d}^2~\Gen}{(\mathrm{d}Q^2)^2}%
\biggl {\vert}_{_{\sQsq}}\hspace*{-6mm}%
=\frac{\rquad}{60}=\frac{a\mu_n }{4M_n^2}%
\left [\frac{4}{\Lambda^2} +  \frac{d}{2M_n^2}\right],
\ee
\be   \label{eq:genderivrat} 
{\cal R} \equiv  \frac{\Gen''(0)}{\Gen'(0)} =
-\frac{1}{10}\frac{\rquad}{\rms} = - \left [\frac{4}{\Lambda^2} +
\frac{d}{2M_n^2} \right] \ .
\ee
The  r.h.s. are obtained by taking the first and second
derivative of the common parametrization of \Gen
\be\label{eq:paramfit}
\Gen(\Qsq) = -\mu_n \cdot {a \tau \over 1 + d \tau}\cdot\Gdipole(\Qsq)
\ee
with \Gdipole(\Qsq) = 1/(1+\Qsq/$\Lambda^2$)$^2$,
 $\Lambda^2$=0.71(GeV/c)$^2$,~and $\tau$=$\Qsq/(4M_n^2)$ (see
 refs.~\cite{galster,plat}), and with $\mu_n$ the neutron magnetic moment. This
 simple parametrization guarantees the proper behaviour of \Gen\ at \Qsq=0
 (zero net charge) and at \Qsq$\to$$\infty$ (quark counting rules).

Similarly, the low momentum expansion of \GCndelta\ can be written as
\be\label{eq:gcndelta}
\GCndeltaQ\ = \Qndelta\ \bigg( 1 - \frac{\rdelta}{6}{\Qsq} + \cdots
\bigg)\ .
\ee
Referring to the connection between \GCndelta\ and \Gen\ (Eq.\ref{eq:gcgenff})
 and equating the respective coefficients of Eqs.(\ref{eq:gcndelta})
 and~(\ref{eq:genexpand}) one finds with the help of
 Eqs.(\ref{eq:genderivone}-\ref{eq:genderivrat})
\bea
\Qndelta\ &=& \frac{\rms}{\sqrt{2}}= \frac{3a}{2\sqrt{2}}\frac{\mu_n}{M_n^2}\\
\rdelta\ &=& \frac{6}{20\sqrt{2}}\frac{\rquad}{\Qndelta}=
\frac{3}{10}\frac{\rquad} {\rms} = -3~{\cal R} \ .
\eea

\begin{figure}[htb]
\begin{center}
\epsfxsize 8.7 true cm
\epsfbox{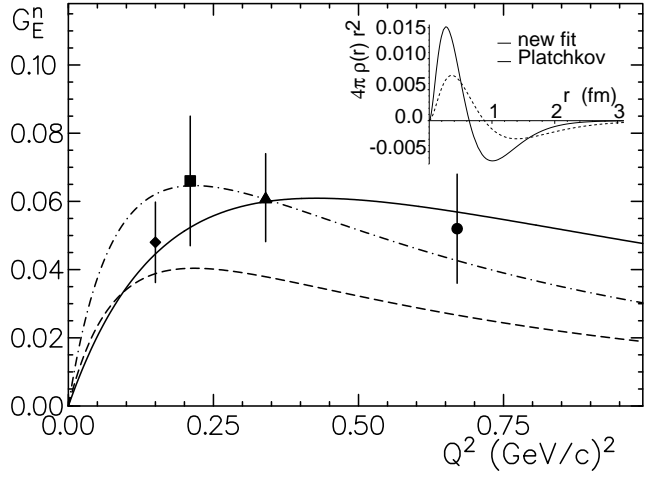}
\vspace{0.3 cm}
\caption{\label{fig:neutff99}
  Recent measurements of \Gen\ are given by the full points (see
  Table~\ref{tab:genexp}).  The Platchkov parametrization~\protect\cite{plat}
  (dashed) is shown together with a fit to the new data without (dot-dashed)
  and with (full line) inclusion of \rms\ from Ref.~\protect\cite{kopecky}.
  The insert shows the neutron charge distributions resulting from the new
  (full line) and the Platchkov (dashed line) parametrizations.}
\end{center}
\end{figure}

{\it New \Gen\ data and fits -- } 
There has recently been increased interest in the neutron electric form factor
 itself as a testing ground for nucleon models. Furthermore, \Gen\ must be
 known accurately in order to analyze for example parity violation and
 \NDelta\ transition experiments.  The previous evaluation by
 Platchkov~\etal\cite{plat} of elastic electron-deuteron scattering data is
 model-dependent because the analysis relies heavily on the deuteron wave
 function when subtracting all magnetic form factors and the charge
 contribution of the proton in order to isolate \Gen\ from the measured cross
 sections. The analysis employing the Paris potential was then considered as
 the most reliable~\cite{plat}, other choices for the $NN$ potential yielded
 values for \Gen\ differing up to 100\%.  The results were parametrised using
 Eq.(\ref{eq:paramfit}) with the two parameters fitted to $a$=1.25 and
 $d$=18.3. An earlier analysis~\cite{galster} obtained the values $a$=1.0 and
 $d$=10.7, respectively.

%
\begin{table}[htb]
\begin{center}
\caption{\label{tab:genexp}
  \Gen\ data resulting from recent experiments are selected and are
  listed with statistical ($\Delta_{stat}$) and total
  ($\Delta_{tot}$=$\Delta_{stat}+\Delta_{syst}$) errors. The \rms\ 
  values from electron-neutron scattering are given in the last line.}
\vspace*{1mm}
\begin{tabular}{ccccr}
\mbox{\rule[-0.2cm]{0.cm}{0.7cm}}      Q$^2$~~$^a$)  &   \Gen\   &
$\Delta_{stat}$ &  $\Delta_{tot}$ & ref. \\
\hline
&&&&\\[-2ex]
     0.15  &     0.0480   &    0.0065 & 0.0118 &\protect\cite{herberg}\\
     0.21  &     0.0660   &    0.0150 & 0.0190 &\protect\cite{paschier}\\
     0.34  &     0.0611   &    0.0069 & 0.0129 &\protect\cite{ostrick}\\
     0.67  &     0.0520   &    0.0110 & 0.0160 &\protect\cite{rohe}\\[1ex]
\hline
\hline
\mbox{\rule[-0.2cm]{0.cm}{0.6cm}} & \rms~$^b$) & $\Delta_{stat}$~$^b$) &
$\Delta_{tot}$~$^b$) & \\
\hline
&&&&\\[-2ex]
     0.0   &     -0.113   &    0.0026 & 0.0060 & \protect\cite{kopecky}\\
\end{tabular}
\end{center}
\vspace*{-4mm}
$^a$)  in units of (GeV/c)$^2$ \hfill $^b$)  in units of fm$^2$ 
\end{table}

In modern double-polarisation experiments, where the ratio of \Gen/\Gmn\ is
 measured in quasifree kinematics, the NN-potential dependence is greatly
 reduced. The remaining uncertainty resides mainly in the corrections for
 final state interaction (FSI) and to a lesser extent for meson exchange
 currents (MEC) between the nucleons.  Recent experiments have been performed
 at Bates, Amsterdam and Mainz by using polarised electrons scattering either
 from unpolarised deuterium and measuring the recoil polarisation of the
 neutron~\cite{herberg,paschier,ostrick,eden}, or by measuring helicity
 dependences on polarised $^3$He~\cite{rohe,meyerhoff,becker,jones}.  Only
 four of these data with sufficient statistical and systematic accuracy have
 been selected (Table~\ref{tab:genexp}) and are displayed in
 Fig.~\ref{fig:neutff99}. A further selection criterium was the coincident
 detection of 
 neutrons and 
 scattered electrons which enables a check on the
 quasifree scattering mechanism.

The three deuteron data~\cite{herberg,paschier,ostrick} rely on corrections
 based on calculations by Arenh\"ovel~\cite{arenhovel}.  Corrections for the
 $^3$He data at \Qsq=0.35~(GeV/c)$^2$~\cite{meyerhoff,becker} are expected to
 be quite large. Therefore, these $^3$He data were not considered in the
 present fits. Note however, that the correction at \Qsq=0.67~(GeV/c)$^2$ is
 estimated to amount to only 10~\%, which has not been applied to the data but
 is included in the systematical error~\cite{rohe}.

Statistical errors are given in column~3 of Table~\ref{tab:genexp}.  The sum
 $\Delta_{tot}$=$\Delta_{stat}$+$\Delta_{syst}$ of statistical and
 systematical errors is shown separately in column~4.  In view of the
 discussion above, we will use $\Delta_{tot}$ in this analysis. The data are
 shown in Fig.~\ref{fig:neutff99} together with the (dashed) curve based on
 the Platchkov parametrization with the Paris potential~\cite{plat}. We note
 that the new data lie significantly above the previous evaluation.

Parameters $a$ and $d$ of Eq.(\ref{eq:paramfit}) have been obtained by a
 ``downhill simplex'' fit using the four data points from quasifree scattering
 and the \rms -value~\cite{kopecky} obtained from thermal neutron electron
 scattering.  The results and their standard deviations due to the fit with
 $\Delta_{tot}$ are compiled in Table~\ref{tab:genfit}, together with the
 total $\chi^2$.  For reference, the parametrizations of
 Galster~\cite{galster} and Platchkov~\cite{plat} are listed in the last two
 lines of Table~\ref{tab:genfit} showing large values for $\chi^2$. The first
 two lines of Table~\ref{tab:genfit} present the fitting results for different
 conditions as marked in the first column.
 From Eq.(\ref{eq:genderivone}) it is evident that parameter $a$ is determined
 exclusively by \rms, which leads to a strong reduction of $\Delta a$ when
 \rms\ is included in the fit.  The result including this constraint (line~2
 of Table~\ref{tab:genfit}) is regarded as the most reliable, considering the
 present status of the data base. It is represented by the full curve in
 Fig.~\ref{fig:neutff99}. The influence of omitting the \rms-datum in the fit
 (line~1 of Table~\ref{tab:genfit}) is shown by the dot-dashed line. From this
 fit a neutron charge radius of \rms=-0.178(27)~fm$^2$ would be deduced which
 is ruled out by the analysis of Ref.~\cite{kopecky}. Despite its large
 uncertainty the parameter $d$ is much smaller than previously
 assumed~\cite{plat} which is reflected by smaller values for \Gen$''(0)$ and
 ${\cal R}$ as shown in the last two columns.

%
\begin{table}[htb]
\begin{center}
\caption{\label{tab:genfit}
  Parameters $a$ and $d$ from fits to the data of Table~\ref{tab:genexp}
  together with $\chi^2$ are shown in the first two lines; values for
  \Gen$''(0)$~(Eq.~\ref{eq:genderivtwo}) and the ratio ${\cal
  R}$~(Eq.~\ref{eq:genderivrat}) are also given.  The
  next three lines contain results where one additional ficticious data
  point with $\Delta_{tot}$=10~\% at \Qsq=0.9~(GeV/c)$^2$ is included in
  the data set for which the assumed \Gen-value is given in col.~1.
  For reference the results using the parameters of
  Refs.~\protect\cite{galster} and~\protect\cite{plat} are shown in the
  last two lines.}
\vspace*{1mm}
\begin{tabular}{l|rrrrr}
comments &
 \multicolumn{1}{c}{a~$\pm ~\Delta$a}  &
   \multicolumn{1}{c}{d~$\pm ~\Delta$d}   &
     \multicolumn{1}{c}{$\chi^2$}  &
       \multicolumn{1}{c}{$\Gen''$ $^a$)} &
         \multicolumn{1}{c}{${\cal R}$ $^b$)} \\
\hline
~$^c$)       & 1.415 (0.213) & 9.06 (3.30) & 0.09 & -8.26 & -10.77 \\
             & 0.898 (0.044) & 2.74 (1.99) & 0.63 & -3.50 &  -7.19 \\
\hline
\Gen=0.04 &  0.903 (0.040) & 4.30 (0.77) & 1.30  & -3.95  & -8.07\\
\Gen=0.05 &  0.898 (0.041) & 2.78 (0.66) & 0.63  & -3.51  & -7.21\\
\Gen=0.06 &  0.894 (0.040) & 1.75 (0.61) & 1.07  & -3.21  & -6.63\\
\hline
ref.~\cite{galster}& 1.00 (~~~-~~~) & 10.70 (~~-~~) & 10.53 &  -6.35 & -11.69\\
ref.~\cite{plat}   & 1.25 (~~~-~~~) & 18.30 (~~-~~) & 62.65 & -10.83 & -16.00\\
\end{tabular}
\end{center}
\vspace*{-4mm}
$^a$) in units of (GeV/c)$^{-4}$    \hfill $^c$) \rms\ datum omitted \\
$^b$) in units of (GeV/c)$^{-2}$    \hfill 
\end{table}

With the present fit the inverse Fourier transformation of Eq.(\ref{eq:trans})
leads to a neutron charge distribution (full line in insert of
Fig.~\ref{fig:neutff99}) quite different from the previous results
(Ref.~\cite{plat}, dashed line), and thus to different moments.  The inner
region of positive contributions is compressed and the zero crossing point is
shifted by about 0.2~fm towards the center, compared to the old distribution.
Note, that the new zero coincides with the value for the quark core radius
$b$ derived from Eq.~(\ref{eq:rnmec}).

A better determination of parameter $d$, which is crucial for the second
 derivative of \Gen, calls for additional data at high \Qsq\ where corrections
 for FSI and MEC are less important. Although the cross sections decrease with
 \Qsq, a statistical error $\Delta_{stat}$=5~\% is attainable and the chosen
 error $\Delta_{tot}$=10~\% seems to be realistic. Simple values for \Gen\
 have been assumed to demonstrate possible variations of $d$. In
 Table~\ref{tab:genfit} we present three more fits each with one fictitious
 data point added at \Qsq=0.9~(GeV/c)$^2$.  The additional datum at
 \Qsq=0.9~(GeV/c)$^2$ reduces $\Delta d$ by a factor 3. Because the highest
 data point at \Qsq=0.6~(GeV/c)$^2$ does not include any FSI correction, a
 further decrease of parameter~$d$ towards a value of 2 seems likely.

\begin{figure}[htb]
\begin{center}
\epsfxsize 13.0 true cm
\epsfbox{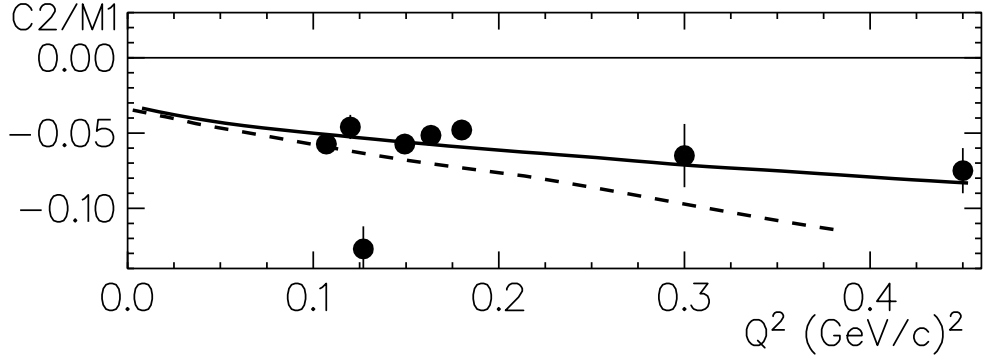}
\vspace{0.3 cm}
\caption{\label{fig:ratio}
  Ratio $C2/M1$ from the present fit to \Gen\ (solid curve) and from
  calculations of Ref.~\protect\cite{Buc00} (dashed curve) in comparison with
  experimental results taken from Refs.~\protect\cite{Bar99,c2old}.}
\end{center}
\end{figure}


{\it Results -- }
The fits of \Gen\ and in particular the first extraction of the fourth moment
 \rquad=-0.32(8)~fm$^4$ permit a prediction for \rdelta=0.84(21)~fm$^2$ where
 the error is determined only through $\Delta d$ (see
 Eq.(\ref{eq:genderivrat})); with the additional sixth data point the error of
 \rdelta\ would be reduced to 0.07~fm$^2$. The variation of $d$ and thus $\cal
 R$ with the sixth data point is indicative of a possible 10~\% change of
 \rdelta.  By use of Eq.(\ref{eq:quadrad}) and a quark core radius $b$=0.6~fm
 we obtain for the charge radius of the light constituent quarks of
 $r^2_{\gamma q}$=0.64~fm$^2$. This value is somewhat larger than the value
 derived from the vector dominance model, which would give $r^2_{\gamma
 q}$=$6/m_{\rho}^2$=0.4~fm$^2$.

%
%

Finally, in the quark model with two-body exchange currents the C2/M1 ratio
 can be expressed as a ratio of the elastic neutron charge and magnetic form
 factors~\cite{Buc00}
\be\label{c2m1}
\frac{C2}{M1} = M_N\  \frac{\sqrt{{\bf q}^2}}{6}\ \frac{\GCndeltaq}{\GMndeltaq}
= \frac{M_N}{2\sqrt{{\bf q}^2}}\ \frac{\Genq}{\Gmnq}  
\ee

Using \Gmn=$\mu_n$\Gdipole\ we compare the predictions of Eq.(\ref{c2m1}) with
 the direct measurements~\cite{Bar99,c2old} of the C2/M1 ratio in pion
 electroproduction.  Starting from the new parametrization of \Gen\ we find
 the ratio in the range of -0.03 to -0.08 (full line in Fig.~\ref{fig:ratio}).
 The data~\cite{Bar99,c2old} in general are in the range of -0.046 to -0.06.
 The agreement is surprising, which hints at a general validity of the
 relation between \Gen\ $\!\!/\!\!$ \Gmn\  
and C2/M1~(Eq.\ref{c2m1}).  A constituent quark
 model calculation~\cite{Buc00} of this ratio (dashed line) comes lower than
 the parametrization.  Differences between this calculation and the C2/M1 data
 may be explained by possible background amplitudes contributing to the
 \NDelta\ transition. Other constituent quark model calculations based on
 one-body currents alone, e.g. Ref.~\cite{capstick}, predict much smaller
 ratios, mostly in the range of -0.005 to -0.02 (see also refs. in
 \cite{Bar99}).

In summary, we have performed a fit to the most reliable data for \Gen\ below
 \Qsq$<$1~(GeV/c)$^2$. We use the obtained parametrization in combination with
 relations originally derived in the constituent quark model to predict the
 quadrupole transition form factor \GCndelta\, and its leading moments
 \Qndelta\ and \rdelta. Eq.~(\ref{eq:gcgenff}) provides a determination of
 C2/M1 through the elastic neutron form factors which agrees well with C2/M1
 data from pion electroproduction experiments.

We gratefully acknowledge the financial support by the 
Deutsche Forschungs\-gemeinschaft (DFG).



\end{document}